\documentclass[10pt]{article}
\usepackage{amsfonts}
\title{Cosmological Implications of a Lorentz Invariance \\Violating O(2) Model}
\author{A. Jahan\\Research Institute for Astronomy and Astrophysics of Maragha
(RIAAM)\\ Maragha, IRAN, P. O. Box: 55134 - 441\\jahan@riaam.ac.ir}
\date{}
\usepackage[english]{babel}
\begin{document}
\maketitle
\begin{abstract}
We derive the free energy of a Lorentz invariance violating (LIV) O(2) model, for which a non-vanishing commutator among it's momenta is supposed. Then we investigate the possible implications of such a matter field
on the time evolution of a flat FRW universe at low and high temperatures.\\\
\
\end{abstract}
\section{Introduction}
Despite the lack of any experimental evidence implying for the breakdown of Lorentz symmetry, there has been speculations on the possible violation from the Lorentz symmetry at high energies, e.g. within the framework of quantum gravity and superstring theories [1]. In a series of recent works some models of fields violating the Lorentz invariance are proposed upon modification of the canonical structure of relativistic fields [2-6]. For example for the O(2) model (two component scalar field model) with
hamiltonian density
\begin{equation}\label{1}
\mathcal{H}=\widehat{\Pi}_i(x)\widehat{{\Pi}}_i(x)+\nabla\phi_i(x)\cdot\nabla{\phi}_i(x),\quad\quad i\in\{1,2\}
\end{equation}
deformation of the canonical structure as
{\setlength\arraycolsep{2pt}
\begin{eqnarray}
[\phi_i(t,\textbf{x}),{\phi}_j(t,\textbf{x}')] &=& 0\\\
[\phi_i(t,\textbf{x}),\widehat{{\Pi}}_j(t,\textbf{x}')]&=&i\delta_{ij}\delta(\textbf{x}-\textbf{x}')\\\
[\widehat{{\Pi}}_i(t,\textbf{x}),\widehat{\Pi}_j(t,\textbf{x}')]&=& i\gamma\epsilon_{ij}\delta(\textbf{x}-\textbf{x}')
\end{eqnarray}}
with $\epsilon_{21}=-\epsilon_{12}=1$, results in a deformed dispersion relation [2]
\begin{equation}\label{6}
E_{1,2}(\textbf{k})=\sqrt{\textbf{k}^2+\frac{\gamma^2}{4}}\pm\frac{\gamma}{2}
\end{equation}
which explicitly violates the relativistic energy-momentum relation $E^2=\textbf{k}^2$. The thermal behavior of a LIV scalar and gauge fields are considered in [7, 8] and possible inflationary scenario for a FRW universe filled with a massless scalar field is examined in [9]. \\\
In this work first we derive the free energy of a LIV O(2) model and then look at its possible effects on the dynamics of a flat FRW universe in high and low temperature limits. It is pointed out that at high temperature with $\gamma\ll 1$ the model evolves as a usual O(2) field. But, compared to the O(2) field with $\gamma=0$, we find a slower expansion at low temperature for the model with $\gamma\neq 0$. Throughout this work we assume $\hbar=c=8\pi G=1$.
\section{Free energy}
The eigenstate $|\,n_{i,\textbf{k}}\rangle$ associated with the field $\phi_i(x)$ satisfies the eigenvalue relations
{\setlength\arraycolsep{2pt}
\begin{eqnarray}\label{1}
H|\,n_{1,\textbf{k}}\rangle &=&E_1(\textbf{k})\Big(n_{1,\textbf{k}}+\frac{1}{2}\Big)|\,n_{1,\textbf{k}}\rangle\\\
H|\,{n}_{2,\textbf{k}}\rangle &=& E_2(\textbf{k})\Big({n}_{2,\textbf{k}}+\frac{1}{2}\Big)|\,{n}_{2,\textbf{k}}\rangle
\end{eqnarray}}
where $n_{i,\textbf{k}}$ stand for the occupation numbers of the \textbf{k}-th mode. So the partition function associated with the Hamiltonian (1) takes the form
\begin{equation}\label{1}
Z=\textrm{Tr}e^{-\beta H}=\prod_{\textbf{k}\neq 0}\frac{1}{1-e^{-\beta E_{2}}}\frac{1}{1-e^{-\beta E_{1}}}
\end{equation}
from which the free energy $\beta F=\ln Z$ is found to be
\begin{equation}
\beta F=V\int\frac{d^dk}{(2\pi)^d}\Big[\ln(1-e^{-\beta E_{2}})+\ln(1-e^{-\beta E_{1}})\Big]
\end{equation}
We first set $\gamma=0$ in (6). Then by means of the expansion
\begin{equation}\label{2}
\ln(1-x)=-\sum_{n=1}^\infty\frac{x^n}{n}
\end{equation}
we find the free energy of a two component massless scalar field in d-dimensions
{\setlength\arraycolsep{2pt}
\begin{eqnarray}
 F&=&-2V\sum^\infty_{n=1}
\int\frac{d^dk}{(2\pi)^d}\frac{e^{-n\beta k}}{\beta n}\\\nonumber
&=&-\frac{4V\Gamma(d)\zeta(d+1)}{(4\pi)^{\frac{d}{2}}\Gamma(\frac{d}{2})}\frac{1}{\beta^{d+1}}
\end{eqnarray}}
For the $\gamma\neq 0$ case the free energy could be written
\begin{equation}
\beta F=-V\sum^\infty_{n=1}\bigg(\frac{e^{\frac{1}{2}n\beta\gamma}}{n}+\frac{e^{-\frac{1}{2}n\beta\gamma}}{n}\bigg)
\int\frac{d^dk}{(2\pi)^d}e^{{-n\beta\sqrt{{\scriptsize{\textbf{k}}}^2+\frac{1}{4}\gamma^2}}}
\end{equation}
On using the formula
\begin{equation}\label{1}
\frac{e^{-xy}}{y}=\frac{1}{\sqrt{2\pi}}\int^\infty_0\frac{dt}{t}e^{-\frac{x^2}{2}t-\frac{y^2}{2t}}
\end{equation}
the integral in (12) can be recast in
\begin{equation}\label{1}
\int\frac{d^dk}{(2\pi)^d}e^{{-n\beta\sqrt{{\scriptsize{\textbf{k}}}^2+\frac{1}{4}\gamma^2}}}=
2n\beta \Big(\frac{\gamma}{4\pi\beta n}\Big)^{\frac{d+1}{2}}
K_{(d+1)/2}\Big(\frac{n\beta\gamma}{2}\Big)
\end{equation}
where $K_{(d+1)/2}$ is the modified bessel function of second kind. Thus the free energy of a LIV scalar field takes the form
\begin{equation}\label{1}
F=-4V\Big(\frac{\gamma}{4\pi\beta}\Big)^{\frac{d+1}{2}}\sum^\infty_{n=1}\frac{1}{n^{(d+1)/2}}\cosh\Big(\frac{n\beta\gamma}{2}\Big)
K_{(d+1)/2}\Big(\frac{n\beta\gamma}{2}\Big)
\end{equation}
By taking into account the Taylor expansion of the Bessel function [10]
\begin{equation}\label{}
K_\nu(z)=\frac{2^{\nu-1}\Gamma(\nu)}{z^\nu}\bigg[1-\frac{z^2}{4(\nu-1)}+O(z^4)\bigg], \quad z\ll 1
\end{equation}
and using $\cosh z=1+\frac{z^2}{2}+O(z^4)$, we get the free energy at high temperature regime
\begin{equation}\label{8}
F=-\frac{2V}{\pi^{\frac{d+1}{2}}}\Gamma\Big(\frac{d+1}{2}\Big)
\bigg[\frac{\zeta({d+1})}{\beta^{d+1}}+\frac{\gamma^2}{8}\frac{d-2}{d-1}\frac{\zeta({d-1})}{\beta^{d-1}}\bigg]
\end{equation}
where $\zeta(a)=\sum_{n=1}^\infty \frac{1}{n^a}$ is the Riemann zeta function. To avoid the singular behavior of (17) at $\nu=1$ we suppose $2<d$. On the other hand the asymptotic form of the Bessel function
\begin{equation}\label{1}
\lim_{z\rightarrow 0}K_\nu(z)=\sqrt{\frac{\pi}{2z}}e^{-z}
\end{equation}
allows one to derive the low temperature limit of the free energy as
\begin{equation}
F=-V\Big(\frac{\gamma}{4\pi}\Big)^{\frac{d}{2}}\zeta\Big(\frac{d}{2}+1\Big)\frac{1}{\beta^{\frac{d}{2}+1}}
\end{equation}
Hence, at low temperature we find for the energy and pressure
{\setlength\arraycolsep{2pt}
\begin{eqnarray}
 \rho&=&\frac{1}{V}\frac{\partial(\beta F)}{\partial\beta}=\frac{d}{2}\Big(\frac{\gamma}{4\pi}\Big)^{\frac{d}{2}}\zeta\Big(\frac{d}{2}+1\Big)\frac{1}{\beta^{\frac{d}{2}+1}} \\\
p&=&-\frac{\partial{F}}{\partial V}=\Big(\frac{\gamma}{4\pi}\Big)^{\frac{d}{2}}\zeta\Big(\frac{d}{2}+1\Big)\frac{1}{\beta^{\frac{d}{2}+1}}
\end{eqnarray}}
which in turn gives rise to the equation of state
\begin{equation}\label{1}
w=\frac{p}{\rho}=\frac{2}{d}
\end{equation}
In high temperature limit we find
{\setlength\arraycolsep{2pt}
\begin{eqnarray}
 \rho&=&\frac{\kappa_0d}{\beta^{d+1}}+\frac{\kappa_2(d-2)}{\beta^{d-1}}\gamma^2 \\\
p&=&\frac{\kappa_0}{\beta^{d+1}}+\frac{\kappa_2}{\beta^{d-1}}\gamma^2
\end{eqnarray}}
Here $\kappa_0$ and $\kappa_2$ are constants depending on $d$ and $\gamma$ (compare (24) with (17) on setting $V=-1$).
\section{Implications for Cosmology}
The Einstein classical action in presence of a finite temperature matter field is
\begin{equation}\label{1}
S=\int d^4x\sqrt g\Big(\frac{R}{2}+\mathcal F\Big)
\end{equation}
where $\mathcal F=F/V$ is the free energy density. The energy-momentum tensor
is defined to be
\begin{equation}\label{1}
T_{\mu\nu}=-\frac{2}{\sqrt{-g}}\frac{\delta}{\delta g^{\mu\nu}}(\sqrt{-g}\mathcal F)
\end{equation}
and has the form
\begin{equation}\label{1}
T_{\mu\nu}=\textrm{diag}(-\rho,p,p,...)
\end{equation}
Therefore from the Einstein equation $R_{\mu\nu}-\frac{1}{2}g_{\mu\nu}=T_{\mu\nu}$, for a isotropic and homogeneous (FRW) space-time with differential line element
\begin{equation}\label{1}
ds^2=-dt^2+\sum_{i=1}^da^2(t)dx_i^2
\end{equation}
one finds the equation of motions
{\setlength\arraycolsep{2pt}
\begin{eqnarray}\label{1}
H^2&=&\frac{2}{d^2-d}\rho\\\
\dot{H}+H^2&=&\frac{1}{d-1}\Big(\frac{2-d}{d}\rho-p\Big)
\end{eqnarray}}
with $H=\frac{\dot a}{a}$ as the Hubble parameter.
\subsection{High Temperature Limit}
By substituting for the pressure and energy form (23) and (24) in (29) and (30) together with assumption $d=3$, equations of motion read
{\setlength\arraycolsep{2pt}
\begin{eqnarray}
H^2&=&\frac{\kappa_0}{\beta^4}+\frac{1}{3}\frac{\kappa_2}{\beta^2}\gamma^2 \\\
\dot{H}+H^2&=&-\frac{\kappa_0}{\beta^4}-\frac{2}{3}\frac{\kappa_2}{\beta^2}\gamma^2
\end{eqnarray}}
Eliminating $\beta$ between (31) and (32) and keeping the terms up to second order in $\gamma$ leads to
\begin{equation}\label{1}
\dot{H}+2H^2+gH=0,\quad\quad g=\frac{\sqrt{5}}{16}\frac{\gamma^2}{\pi}
\end{equation}
where we have used $\kappa_0=\frac{\pi^2}{45}$ and $\kappa_2=\frac{1}{48}$ in $d=3$ dimensions. On using $\dot H+H^2=\frac{\ddot a}{a}$ and $H=\frac{\dot a}{a}$, (33) simplifies to
\begin{equation}\label{1}
\frac{d}{dt}\Big(\dot a a+\frac{g}{2}a^2\Big)=0
\end{equation}
So by demanding $a(0)=0$ one arrives at
\begin{equation}\label{1}
a(t)=a(t_0)\sqrt{\frac{t}{t_0}}\Big[1+\frac{g}{4}(t^2-t^2_0)+O(g^2)\Big]
\end{equation}
In $\gamma\rightarrow 0$ limit we are left with the well-known results for a radiation dominated universe, i.e.
{\setlength\arraycolsep{2pt}
\begin{eqnarray}\label{1}
H&=&\frac{1}{2t}\\\
a&=& a(t_0)\sqrt{\frac{t}{t_0}}
\end{eqnarray}}
So as equation (35) implies, at high temperature for $\gamma\ll1$ the correction due to the break down of the Lorentz symmetry is negligible. This could be understood from the right hand side of (31) where the dominant term is proportional to $\frac{1}{\beta^4}$
\subsection{Low Temperature Limit}
 At low temperatures the equation of state is $w=\frac{2}{d}$. Thus from (22), (29) and (30), and by the use of anzats $a\sim t^\alpha$ and $\rho\sim t^r$ one easily obtains the scale factor and hubble parameters
{\setlength\arraycolsep{2pt}
\begin{eqnarray}
  a &=&a(t_0) \Big(\frac{t}{t_0}\Big)^\frac{2}{d+2} \\
H&=& \frac{2}{d+2}\frac{1}{t}
\end{eqnarray}}
Hence in $d=3$ we have
\begin{equation}\label{1}
a(t) \sim \left\{ \begin{array}{ll}
 t^{\frac{2}{5}} \quad& 1\ll\gamma\beta\\
\sqrt{\frac{t}{t_0}}\Big[1+\frac{g}{4}(t^2-t^2_0)\Big]\quad& \gamma\beta\ll1
\end{array} \right.
\end{equation}
Therefore, at low temperature a d-dimensional universe with LIV scalar field expands as a (d+1)-dimensional universe with a standard matter content for which the Hubble parameter is found to be $H=\frac{2}{d+1}\frac{1}{t} $.

\section{Conclusions}
We derived the free energy of a Lorentz invariance violating O(2) model and considered the dynamics of a FRW space-time filled with such a matter field. We calculated the time evolution of the scale factor up to second order in deformation parameter in high temperature limit. It is pointed out that at low temperature, an O(2) model violating the Lorentz symmetry expands slower than a model with usual standard canonical structure.

\end{document}